\documentclass[11pt]{article}
\usepackage{amssymb}
\usepackage{colortbl}
\usepackage{amsfonts,amsmath, longtable}

\newcommand{\tr}{{\rm tr}}
\newcommand{\ti}[1]{\tilde{#1}}

\newcommand{\mL}{{\mathcal L}}

\newcommand{\vf}{\varphi}

\newcommand{\om}{\omega}

\newcommand{\bfq}{{\bf{q}}}

\newcommand{\bfp}{{\bf{p}}}

\def\beq{\begin{equation}}
\def\eq{\end{equation}}

\newcommand{\mat}[4]{\left(\begin{array}{cc}{#1}&{#2}\\ \ \\{#3}&{#4}
\end{array}\right)}

\newcommand{\mats}[4]{\left(\begin{array}{cc}{#1}&{#2}\\ {#3}&{#4}
\end{array}\right)}

\def\res{\mathop{\hbox{Res}}\limits}

\begin{document}

\setcounter{page}{1}

\begin{center}


\vspace{0mm}

{\Large{\bf  Integrable open elliptic Toda chain with boundaries }}





 \vspace{12mm}

 {\Large {Andrei Zotov}}

  \vspace{8mm}



 {\em Steklov Mathematical Institute of Russian
Academy of Sciences,\\ 8 Gubkina St., Moscow 119991, Russia}

{\em Institute for Theoretical and Mathematical Physics,\\
Lomonosov Moscow State University, Moscow, 119991, Russia}


 {\small\rm {E-mail: zotov@mi-ras.ru}}

\end{center}

\vspace{0mm}

\begin{abstract}
In this letter we discuss the classical integrable elliptic Toda chain proposed by I. Krichever.
Our goal is to construct an open elliptic Toda chain with boundary terms.
This is achieved using the factorized form of the Lax matrix and gauge equivalence with the XYZ chain.
\end{abstract}

%

{\small{ \tableofcontents }}

\parskip 5pt plus 1pt   \jot = 1.5ex

\section{Introduction}
The elliptic closed Toda chain was introduced by I. Krichever \cite{KrToda}. It is a classical integrable
system defined by the following Hamiltonian:
  \beq\label{e01}
  \begin{array}{c}
  \displaystyle{
H^{\hbox{\tiny{eToda}}}=-\frac12\sum\limits_{a=1}^n\Big(\log\frac{1}{\sinh^2(\bfp_a/2c)}+
\log\Big(\wp(\bfq_{a-1}-\bfq_a)-\wp(\bfq_{a-1}+\bfq_a)\Big)\Big)\,,
 }
 \end{array}
 \eq
where $\bfp_a$, $\bfq_a$, $a=1,...,n$ is a set of canonically conjugated variables (that is $\{\bfp_a,\bfq_b\}=\delta_{ab}$)
and $\wp(x)$ is the Weierstrass elliptic function on elliptic curve with moduli $\tau$, ${\rm Im}(\tau)>0$.
We assume here that numeration of particles is modulo $n$: $\bfq_{n+1}=\bfq_1$,
$\bfq_0=\bfq_n$. This model is a particular case of a more general elliptic Ruijsenaars-Toda chain
\cite{Adler}, see also \cite{AdlerShabat,Suris,Yamilov}.

It was recently shown in \cite{MuZ2} that the elliptic Toda chain is a particular case of the Ruijsenaars
chain introduced previously in \cite{ZZ}. The classical $r$-matrix structure for the
Ruijsenaars chain was derived in \cite{MuZ1} and then adopted to the elliptic Toda chain in \cite{MuZ2}.
Another important observation of \cite{MuZ2} and \cite{ZZ} is that the elliptic Toda chain
is gauge equivalent to the discrete Landau-Lifshitz model (or the classical XYZ chain) with some special
restrictions. In the next two sections we briefly recall the main results on this topic.

The purpose of the paper is to derive an open elliptic Toda chain with boundary terms using
the gauge equivalence mentioned above. Our construction is based on
the standard description of the classical transfer-matrix
for XYZ chain with boundary \cite{Skl-refl}. Using the factorized form for the Lax matrices we
represent the transfer-matrix as the one describing the open Toda chain with the gauge transformed $K$-matrices.
This allows to compute the Hamiltonian, and the integrability follows from integrability of
the XYZ chain with boundaries.

The paper is organized as follows. In the next two sections, we recall the results of our previous paper \cite{MuZ2}. In particular, we describe the Lax representation for the elliptic Toda chain and the associated gauge transformation relating the elliptic Toda chain to the closed XYZ chain.
Section \ref{sec4} is devoted to a brief review of Sklyanin's construction \cite{Skl-refl} for classical integrable chains based on boundary $K$-matrices, which are solutions of the classical reflection equation.
The main results of this paper are presented in the last two sections.
In Section \ref{sec5}, we perform the gauge transformation for the open XYZ chain and compute the corresponding gauge-transformed $K$-matrices.
In Section \ref{sec6}, we derive the Hamiltonian of the elliptic Toda chain. The interaction terms for the internal sites of the chain are calculated in a manner similar to that used for the closed chain in \cite{MuZ2}. The boundary terms are also evaluated, yielding the final result.

\section{Lax matrices in factorized form}\label{sec2} Introduce the following matrix:
\beq\label{e02}
  \begin{array}{c}
  \displaystyle{
g(z,\bfq_a)=\frac{1}{\theta_1(2\bfq_a|\tau)}
\mat{\theta_3(z-2\bfq_a|2\tau)}{-\theta_3(z+2\bfq_a|2\tau)}{-\theta_2(z-2\bfq_a|2\tau)}{\theta_2(z+2\bfq_a|2\tau)}\,.
}
 \end{array}
 \eq
This is an intertwining matrix entering the IRF-Vertex correspondence \cite{Baxter2} in quantum statistical models.
Its geometric interpretation and applications to interrelations between classical integrable systems can be found in
\cite{LOZ,VZ,AtZ}.
We use the standard notations for Jacobi theta functions.
From here on it is assumed that all elliptic functions depend on $\tau$ unless otherwise stated explicitly
(like in (\ref{e02})).
Let us introduce a notation for a set of $2\times 2$ matrices:
\beq\label{e03}
  \begin{array}{c}
  \displaystyle{
L^{[k,m]}(z)=-g^{-1}(z,\bfq_k)g(z,\bfq_m)e^{P^m/2c}\,,\quad k,m=1,...,n\,,
}
 \end{array}
 \eq
where $P^m$ are diagonal $2\times 2$ matrices of particles momenta:
\beq\label{e031}
  \begin{array}{c}
  \displaystyle{
P^m={\rm diag}(\bfp_m,-\bfp_m)\,,\quad m=1,...,n\,.
}
 \end{array}
 \eq
Direct calculation yields expression for $L^{[k,m]}(z)$
\beq\label{e04}
  \begin{array}{c}
  \displaystyle{
L^{[k,m]}(z)=\mat{\phi(z,\bfq_{k}-\bfq_m)b^{[k,m]}_1}{\phi(z,\bfq_{k}+\bfq_m)b^{[k,m]}_2}
{\phi(z,-\bfq_{k}-\bfq_m)b^{[k,m]}_1}{\phi(z,-\bfq_{k}+\bfq_m)b^{[k,m]}_2}\,,
}
 \end{array}
 \eq
where $\phi(z,x)$ is the elliptic Kronecker function
\beq\label{e041}
  \begin{array}{c}
  \displaystyle{
\phi(z,u)=\frac{\theta_1'(0)\theta_1(z+u)}{\theta_1(z)\theta_1(u)}
}
 \end{array}
 \eq
and $b^{[k,m]}_1$, $b^{[k,m]}_2$ are the following functions:
\beq\label{e06}
  \begin{array}{c}
  \displaystyle{
b^{[k,m]}_1=\frac{\theta_1(\bfq_m-\bfq_{k})\theta_1(\bfq_m+\bfq_{k})}{\theta_1'(0)\theta_1(2\bfq_m)}\,
e^{{\bfp_m}/{2c}}\,,
}
\\ \ \\
  \displaystyle{
b^{[k,m]}_2=-\frac{\theta_1(\bfq_m-\bfq_{k})\theta_1(\bfq_m+\bfq_{k})}{\theta_1'(0)\theta_1(2\bfq_m)}\,
e^{-{\bfp_m}/{2c}}\,.
}
 \end{array}
 \eq
The Lax matrices for the elliptic Toda chain can be written in several different ways.
First possibility is to use $L^{[a-1,a]}(z)$, $a=1,...,n$ from the definition (\ref{e04}). The monodromy matrix $T^{\hbox{\tiny{eToda}}}(z)$ takes the form \cite{MuZ2}:
\beq\label{e07}
  \begin{array}{c}
  \displaystyle{
T^{\hbox{\tiny{eToda}}}(z)=L^{[n,1]}(z)L^{[1,2]}(z)L^{[2,3]}(z)\dots L^{[n-1,n]}(z)\,.
}
 \end{array}
 \eq
Here $z$ is the spectral parameter. Being written in this from, the elliptic Toda chain is a particular case
of ${\rm GL}_N$ elliptic Ruijsenaars chain on $n$ sites introduced
 in \cite{ZZ}. Its classical $r$-matrix structure was described in \cite{MuZ1}. More precisely,
${\rm GL}_2$ elliptic Ruijsenaars chain depends on the additional deformation parameter $\eta$, and
the elliptic Toda chain arises when $\eta=0$, while for $\eta\neq 0$ it reproduces another
known integrable model -- the elliptic Ruijsenaars-Toda chain\footnote{In fact, the elliptic Ruijsenaars-Toda chain
contains $n$ parameters $\eta_a$, $a=1,...,n$. That is, the elliptic Ruijsenaars-Toda chain with $\eta_a=0$ for all $a$
yields the elliptic Toda chain, and the elliptic Ruijsenaars-Toda chain with pairwise equal $\eta_a=\eta_b$ corresponds
to ${\rm GL}_2$ elliptic Ruijsenaars chain (with reduction to the center of mass frames at each site).
Details of interrelations between models can be found in
\cite{MuZ2}.} described in \cite{Adler}.

Alternatively, one can use the modified Lax representation:
\beq\label{e08}
  \begin{array}{c}
  \displaystyle{
{\bf T}^{\hbox{\tiny{eToda}}}(z)={\bf L}^{1}(z){\bf L}^{2}(z)\dots {\bf L}^{n}(z)
}
 \end{array}
 \eq
with
\beq\label{e081}
  \begin{array}{c}
  \displaystyle{
{\bf L}^{a}(z)=\frac{1}{b^{[a-1,a]}_1+b^{[a-1,a]}_2}\, L^{[a-1,a]}(z)\,.
}
 \end{array}
 \eq
Explicitly, we have
\beq\label{e10}
  \begin{array}{c}
  \displaystyle{
{\bf L}^a(z)=\frac12\mat{\phi(z,\bfq_{a-1}-\bfq_a)
\displaystyle{\frac{e^{\bfp_a/2c}}{\sinh(\bfp_a/2c)}}}
{-\phi(z,\bfq_{a-1}+\bfq_a)\displaystyle{\frac{e^{-\bfp_a/2c}}{\sinh(\bfp_a/2c)}}}
{\phi(z,-\bfq_{a-1}-\bfq_a)\displaystyle{\frac{e^{\bfp_a/2c}}{\sinh(\bfp_a/2c)}}}
{-\phi(z,-\bfq_{a-1}+\bfq_a)\displaystyle{\frac{e^{-\bfp_a/2c}}{\sinh(\bfp_a/2c)}}}\,.
}
 \end{array}
 \eq
The classical transfer-matrices in these two descriptions are related as ${\bf T}^{\hbox{\tiny{eToda}}}(z)=2^{-n}e^{-H^{\hbox{\tiny{eToda}}}}{T}^{\hbox{\tiny{eToda}}}(z)$.
 Using the identity
\beq\label{e101}
  \begin{array}{c}
  \displaystyle{
\phi(z,u)\phi(z,-u)=\wp(z)-\wp(u)
}
 \end{array}
 \eq
for the function (\ref{e041})  we have
\beq\label{e11}
  \begin{array}{c}
  \displaystyle{
\det{\bf L}^a(z)=\frac14\frac{1}{\sinh^2(\bfp_a/2c)}\Big(\wp(\bfq_{a-1}-\bfq_a)-\wp(\bfq_{a-1}+\bfq_a)\Big)
}
 \end{array}
 \eq
for ${\bf L}^a(z)$ (\ref{e10}). Notice that it is independent of spectral parameter $z$.
Then it is easy to see that the expression
\beq\label{e111}
  \begin{array}{c}
  \displaystyle{
\log\det{\bf T}^{\hbox{\tiny{eToda}}}(z)=\sum\limits_{a=1}^n \log\det{\bf L}^a(z)
}
 \end{array}
 \eq
  provides\footnote{More precisely, it is $-2H^{\hbox{\tiny{eToda}}}$ up to an additive
constant coming from the log(1/4) terms} the original Hamiltonian (\ref{e01}).

Details of the above results can be found in \cite{ZZ,MuZ2}. Let us also mention that we use the factorization formulae
of type (\ref{e03}). Originally, this type representation of the Lax matrices was observed by K. Hasegawa in \cite{Has}
at quantum level for the Ruijsenaars-Schneider model. See also a review \cite{VZ} and references therein.

\section{Gauge equivalence with XYZ chain}\label{sec3}
Plugging the factorized representation (\ref{e03})
into (\ref{e07}) gives
\beq\label{e12}
  \begin{array}{c}
  \displaystyle{
T^{\hbox{\tiny{eToda}}}(z)=
 }
 \\ \ \\
  \displaystyle{
=(-1)^n g^{-1}(z,\bfq_n)g(z,\bfq_1)e^{P^1/2c}
g^{-1}(z,\bfq_1)g(z,\bfq_2)e^{P^2/2c}...g^{-1}(z,\bfq_{n-1})g(z,\bfq_n)e^{P^n/2c}\,.
}
 \end{array}
 \eq
Therefore, the gauge transformed monodromy matrix
\beq\label{e131}
  \begin{array}{c}
  \displaystyle{
{\mathcal T}(z)=g(z,\bfq_n)T^{\hbox{\tiny{eToda}}}(z) g^{-1}(z,\bfq_n)
}
 \end{array}
 \eq
takes the form
\beq\label{e13}
  \begin{array}{c}
  \displaystyle{
{\mathcal T}(z)=(-1)^n \mL^1(z)\mL^2(z)\dots \mL^n(z)
}
 \end{array}
 \eq
with the Lax matrices
\beq\label{e132}
  \begin{array}{c}
  \displaystyle{
\mL^a(z)=g(z,\bfq_a)e^{P^a/2c}g^{-1}(z,\bfq_a)\,,\quad a=1,...,n\,.
}
 \end{array}
 \eq
It was explained in \cite{MuZ2} that $\mL^a(z)$ are the Lax matrices of the XYZ chain,
and ${\mathcal T}(z)$ is its standard monodromy matrix \cite{Skl4,FT}.
Namely, the matrices $\mL^a(z)$ are represented in the form:
 \beq\label{e14}
 \begin{array}{c}
  \displaystyle{
\mL^a(z)=\sigma_0{\bf S}^a_0+\sum\limits_{k=1}^3\sigma_k\vf_k(z){\bf S}^a_k\,,
 }
 \end{array}
 \eq
 where $\sigma_a$ are the Pauli matrices
 \beq\label{e15}
 \begin{array}{c}
  \displaystyle{
 \sigma_0=\mats{1}{0}{0}{1}\,,\quad
  \sigma_1=\mats{0}{1}{1}{0}\,,\quad
  \sigma_2=\mats{0}{-\imath}{\imath}{0}\,,\quad
   \sigma_3=\mats{1}{0}{0}{-1}
  }
 \end{array}
\eq
 and
 \beq\label{e16}
 \begin{array}{c}
  \displaystyle{
  \vf_{1}(z)=\frac{\theta_1'(0)\theta_{4}(z)}{\theta_{1}(z)\theta_{4}(0)}\,,\qquad
  \vf_{2}(z)=\frac{\theta_1'(0)\theta_{3}(z)}{\theta_{1}(z)\theta_{3}(0)}\,,\qquad
   \vf_{3}(z)=\frac{\theta_1'(0)\theta_{2}(z)}{\theta_{1}(z)\theta_{2}(0)}\,,
  }
 \end{array}
\eq
 or, in the short form
 \beq\label{e161}
 \begin{array}{c}
  \displaystyle{
 \vf_{k}(z)=\frac{\theta_1'(0)\theta_{5-k}(z)}{\theta_{1}(z)\theta_{5-k}(0)}\,,\quad k=1,2,3\,.
   }
 \end{array}
\eq
 The variables ${\bf S}^a_k$ entering (\ref{e14}) are expressed through $\bfp_a$ and $\bfq_a$ as follows:
\beq\label{e17}
  \begin{array}{c}
  \displaystyle{
{\bf S}_0^a=\frac12\Big(e^{\bfp_a/2c}+
e^{-\bfp_a/2c}\Big)\,,
}
\\ \ \\
  \displaystyle{
{\bf S}_k^a=-\frac{\imath^{\delta_{k,2}}}2\frac{\theta_{5-k}(0)}{\theta_1'(0)}
\Big(\frac{\theta_{5-k}(2\bfq_a)}{\theta_1(2\bfq_a)}\,e^{\bfp_a/2c}-
\frac{\theta_{5-k}(2\bfq_a)}{\theta_1(2\bfq_a)}\,e^{-\bfp_a/2c}\Big)\,,\quad k=1,2,3
}
 \end{array}
 \eq
for all $a=1,...,n$. This is a particular case of a more general formulae\footnote{In the continuous ($n\to\infty$) and non-relativistic ($c\to\infty$) limit the XYZ chain turns into 1+1 XYZ Landau-Lifshitz model. It is gauge equivalent to 1+1 field analogue of 2-body
Calogero-Moser model \cite{LOZ,AtZ}. In this respect, the elliptic Toda chain in the continuous non-relativistic limit
provides this 1+1 field analogue of 2-body
Calogero-Moser model with zero coupling constant.}:
 \beq\label{e170}
 \begin{array}{c}
  \displaystyle{
\mL^a(z,\eta)=\sigma_0{\bf S}^a_0(\eta)+\sum\limits_{k=1}^3\sigma_k\vf_k(z){\bf S}^a_k(\eta)\,,
 }
 \end{array}
 \eq
\beq\label{e171}
  \begin{array}{c}
  \displaystyle{
{\bf S}_0^a(\eta)=\frac12\Big(\frac{\theta_1(2\bfq_a-\eta)}{\theta_1(2\bfq_a)}\,e^{\bfp_a/2c}+
\frac{\theta_1(2\bfq_a+\eta)}{\theta_1(2\bfq_a)}\,e^{-\bfp_a/2c}\Big)\,,
}
\\ \ \\
  \displaystyle{
{\bf S}_k^a(\eta)=-\frac{\imath^{\delta_{k,2}}}2\frac{\theta_{5-k}(0)}{\theta_1'(0)}
\Big(\frac{\theta_{5-k}(2\bfq_a-\eta)}{\theta_1(2\bfq_a)}\,e^{\bfp_a/2c}-
\frac{\theta_{5-k}(2\bfq_a+\eta)}{\theta_1(2\bfq_a)}\,e^{-\bfp_a/2c}\Big)\,,\quad k=1,2,3\,.
}
 \end{array}
 \eq
The Poisson brackets between ${\bf S}_k^a(\eta)$ computed through $\{\bfp_a,\bfq_b\}=\delta_{ab}$
reproduce the classical Sklyanin algebra \cite{Skl2}:
\beq\label{e19}
  \begin{array}{c}
  \displaystyle{
c\{{\bf S}^a_i(\eta),{\bf S}^a_j(\eta)\}=\imath\varepsilon_{ijk} {\bf S}^a_0(\eta) {\bf S}^a_k(\eta)\,,
}
\\ \ \\
  \displaystyle{
c\{{\bf S}^a_0(\eta),{\bf S}^a_i(\eta)\}=
\imath\varepsilon_{ijk} {\bf S}^a_j(\eta) {\bf S}^a_k(\eta)\big(\wp(\om_j)-\wp(\om_k)\big)\,,
}
 \end{array}
 \eq
where
the half-periods $\om_j$ (of the underlying elliptic curve with moduli $\tau$)
 \beq\label{e191}
 \begin{array}{c}
  \displaystyle{
 \om_1=\frac{\tau}{2}\,,\quad
 \om_2=\frac{1+\tau}{2}\,,\quad
 \om_3=\frac{1}{2}
  }
 \end{array}
\eq
are numerated according to numeration of the Pauli matrices.
The expressions ${\bf S}^a_k(\eta)$ (\ref{e171}) are the classical analogues of the
representation of quantum Sklyanin algebra through difference operators (the shift operators
become exponents of momenta in classical mechanics).
The limit $\eta\to 0$ is regular for the functions ${\bf S}^a_k(\eta)$ (\ref{e171}),
and the structure constants of the Sklyanin algebra (\ref{e19}) are independent of $\eta$.
Therefore, expressions ${\bf S}^a_k$ (\ref{e17}) have the same Poisson brackets:
\beq\label{e192}
  \begin{array}{c}
  \displaystyle{
c\{{\bf S}^a_i,{\bf S}^a_j\}=\imath\varepsilon_{ijk} {\bf S}^a_0 {\bf S}^a_k\,,
}
\\ \ \\
  \displaystyle{
c\{{\bf S}^a_0,{\bf S}^a_i\}=
\imath\varepsilon_{ijk} {\bf S}^a_j {\bf S}^a_k\big(\wp(\om_j)-\wp(\om_k)\big)\,.
}
 \end{array}
 \eq
See details in \cite{MuZ2} and \cite{ZZ}.

The Sklyanin algebra (\ref{e19}) is generated by the classical quadratic exchange relation
for the Lax matrices (\ref{e14})
 \beq\label{e20}
 \begin{array}{c}
  \displaystyle{
 \{\mL_1^{a}(z,\eta),\mL_2^{b}(w,\eta)\}
 =\frac{\delta^{ab}}{c}\,[\mL_1^{a}(z,\eta)\mL_2^{a}(w,\eta),r_{12}(z-w)]
 }
 \end{array}
 \eq
 with the elliptic $r$-matrix
 \beq\label{e201}
 \begin{array}{c}
  \displaystyle{
 r_{12}(z-w)=\frac12\sum\limits_{k=1}^3\vf_k(z-w)\sigma_k\otimes\sigma_k\,.
 }
 \end{array}
 \eq
 The center of the algebra (\ref{e20}) (and therefore, center of (\ref{e171}) or (\ref{e19}))
 is generated by
\beq\label{e202}
  \begin{array}{c}
  \displaystyle{
\det {\mL}^a(z,\eta)={\bf C}_2^a(\eta)-\wp(z){\bf C}_1^a(\eta)
}
 \end{array}
 \eq
with the Casimir functions
\beq\label{e203}
  \begin{array}{c}
  \displaystyle{
{\bf C}_1^a(\eta)=\big({\bf S}^a_1(\eta)\big)^2+\big({\bf S}^a_2(\eta)\big)^2+\big({\bf S}^a_3(\eta)\big)^2\,,
}
\\ \ \\
  \displaystyle{
{\bf C}_2^a(\eta)=\big({\bf S}^a_0(\eta)\big)^2+\sum\limits_{k=1}^3 \big({\bf S}^a_k(\eta)\big)^2\wp(\om_k)\,.
}
 \end{array}
 \eq
The representation (\ref{e171}) yields
\beq\label{e204}
  \begin{array}{c}
  \displaystyle{
{\bf C}_1^a(\eta)=\Big(\frac{\theta_1(\eta)}{\theta_1'(0)}\Big)^2\,,
\qquad
{\bf C}_2^a(\eta)=\wp(\eta)\Big(\frac{\theta_1(\eta)}{\theta_1'(0)}\Big)^2\,.
}
 \end{array}
 \eq
The representation (\ref{e17}) appears from (\ref{e171}) in the limit $\eta\to 0$. Then
for ${\bf C}_1^a=({\bf S}^a_1)^2+({\bf S}^a_2)^2+({\bf S}^a_3)^2$ and
${\bf C}_2^a=({\bf S}^a_0)^2+\sum\limits_{k=1}^3 ({\bf S}^a_k)^2\wp(\om_k)$ defined through
the functions (\ref{e17}) we have
\beq\label{e22}
  \begin{array}{c}
  \displaystyle{
{\bf C}_1^a=0\,,
\qquad
{\bf C}_2^a=1\,.
}
 \end{array}
 \eq
Then from (\ref{e202}) we also conclude that for the Lax matrices  ${\mL}^a(z)$ (\ref{e14})
\beq\label{e221}
  \begin{array}{c}
  \displaystyle{
\det {\mL}^a(z)=\det {\mL}^a(z,\eta)\big|_{\eta=0}=1\,.
}
 \end{array}
 \eq

\section{XYZ chain with boundaries}\label{sec4}
Let us recall main steps for construction of integrable chains with boundaries following \cite{Skl-refl}.
Consider the monodromy matrix:
\beq\label{e40}
  \begin{array}{c}
  \displaystyle{
{\mathcal T}^{\hbox{\tiny{openXYZ}}}(z)=
K^+(z){\mathcal T}(z)K^-(z){\mathcal T}^{-1}(-z)\,,
}
 \end{array}
 \eq
where ${\mathcal T}(z)$ (\ref{e13}) satisfies the quadratic classical exchange relation
\beq\label{e41}
  \begin{array}{c}
  \displaystyle{
  c\{{\mathcal T}_1(z),{\mathcal T}_2(w)\}=[{\mathcal T}_1(z){\mathcal T}_2(w),r_{12}(z-w)]
  }
 \end{array}
 \eq
due to each Lax matrix $\mL^a(z)$ satisfies the same quadratic relation (\ref{e20}),
 and $K^\pm(z)$ are $2\times 2$ matrices solving the classical quadratic reflection equation
\beq\label{e42}
\displaystyle{
[K^\pm_1(z)K^\pm_2(w),r _{12}(z-w)]+\,K^\pm_2(w)r_{12}(z+w)K^\pm_1(z)
-\,K^\pm_1(z)r_{12}(z+w)K^\pm_2(w)=0
}
\eq
with the same elliptic $r$-matrix from (\ref{e20}). Then the classical transfer-matrix
$\tr{\mathcal T}^{\hbox{\tiny{openXYZ}}}(z)$ is a generating function of Poisson commuting Hamiltonians since
\beq\label{e43}
  \begin{array}{c}
  \displaystyle{
  \{ \tr{\mathcal T}^{\hbox{\tiny{openXYZ}}}(z), \tr{\mathcal T}^{\hbox{\tiny{openXYZ}}}(w) \}=0\,.
  }
 \end{array}
 \eq
In order to compute ${\mathcal T}^{-1}(-z)=(-1)^n (\mL^n(-z))^{-1}(\mL^{n-1}(-z))^{-1}\dots (\mL^1(-z))^{-1}$ (\ref{e13}) entering (\ref{e40}) we need to find $({\mL}^a(-z))^{-1}$.
For any matrix of the form (\ref{e14}) we have
\beq\label{e431}
  \begin{array}{c}
  \displaystyle{
 \Big({\mL}^a(z)\Big)^{-1}=\frac{1}{\det({\mL}^a(z))}\Big(\sigma_0{\bf S}^a_0-\sum\limits_{k=1}^3\sigma_k\vf_k(z){\bf S}^a_k\Big)\,.
  }
 \end{array}
 \eq
The functions $\vf_k(z)$ (\ref{e16}) are odd in $z$:
\beq\label{e4311}
  \begin{array}{c}
  \displaystyle{
 \vf_k(-z)=-\vf_k(z)\,.
  }
 \end{array}
 \eq
Therefore,
\beq\label{e432}
  \begin{array}{c}
  \displaystyle{
 \Big({\mL}^a(-z)\Big)^{-1}=\frac{1}{\det({\mL}^a(-z))}\Big(\sigma_0{\bf S}^a_0+\sum\limits_{k=1}^3\sigma_k\vf_k(z){\bf S}^a_k\Big)={\mL}^a(z)\,,
  }
 \end{array}
 \eq
where we also used $\det({\mL}^a(-z))=\det({\mL}^a(z))=1$ because of (\ref{e221}) in our case. Thus,
\beq\label{e433}
  \begin{array}{c}
  \displaystyle{
{\mathcal T}^{-1}(-z)=(-1)^n \mL^n(z)\mL^{n-1}(z)\dots \mL^1(z)\,,
  }
 \end{array}
 \eq
and
\beq\label{e45}
  \begin{array}{c}
  \displaystyle{
{\mathcal T}^{\hbox{\tiny{openXYZ}}}(z)=
K^+(z)\mL^1(z)\mL^2(z)\dots \mL^n(z)K^-(z)\mL^n(z)\mL^{n-1}(z)\dots \mL^1(z)\,.
}
 \end{array}
 \eq

\section{Gauge transformed $K$-matrices}\label{sec5}
Plugging the factorized representation $\mL^k(z)=g(z,\bfq_k)e^{P^k/2c}g^{-1}(z,\bfq_k)$ (\ref{e13}) into
 (\ref{e45}) we get
\beq\label{e46}
  \begin{array}{c}
  \displaystyle{
{\mathcal T}^{\hbox{\tiny{openXYZ}}}(z)=
}
\\ \ \\
  \displaystyle{
=K^+(z)g(z,\bfq_1)e^{P^1/2c}g^{-1}(z,\bfq_1)\dots g(z,\bfq_n)e^{P^n/2c}g^{-1}(z,\bfq_n)\times
}
\\ \ \\
  \displaystyle{
\times K^-(z)
g(z,\bfq_n)e^{P^n/2c}g^{-1}(z,\bfq_n)\dots g(z,\bfq_1)e^{P^1/2c}g^{-1}(z,\bfq_1)\,.
}
 \end{array}
 \eq
Consider the gauge transformed monodromy matrix:
\beq\label{e470}
  \begin{array}{c}
  \displaystyle{
{\mathcal T}^{\hbox{\tiny{openToda}}}(z)=g^{-1}(z,\bfq_1){\mathcal T}^{\hbox{\tiny{openXYZ}}}(z)g(z,\bfq_1)\,.
}
 \end{array}
 \eq
Using the notation (\ref{e03}) it is written in the form:
\beq\label{e47}
  \begin{array}{c}
  \displaystyle{
{\mathcal T}^{\hbox{\tiny{openToda}}}(z)=
}
\\ \ \\
  \displaystyle{
={\ti K}^+(z)e^{P^1/2c}L^{[1,2]}(z)\dots
L^{[n-1,n]}(z){\ti K}^-(z)e^{P^n/2c}L^{[n,n-1]}(z)\dots L^{[2,1]}(z)\,,
}
 \end{array}
 \eq
where
\beq\label{e48}
  \begin{array}{c}
  \displaystyle{
{\ti K}^+(z)=g^{-1}(z,\bfq_1){K}^+(z)g(z,\bfq_1)\,,
}
\\ \ \\
  \displaystyle{
 {\ti K}^-(z)=g^{-1}(z,\bfq_n){K}^-(z)g(z,\bfq_n)\,.
}
 \end{array}
 \eq
The elliptic $K$-matrix solving the reflection equation (\ref{e42}) is known from \cite{IK}.
We write it in the following form:
 \beq\label{e49}
 \begin{array}{c}
  \displaystyle{
K^\pm(z)=\sigma_0\nu_0^\pm+\frac{\nu^\pm_1}{c_1(\tau)\vf_1(z)}\,\sigma_1
-\imath\frac{\nu^\pm_2}{c_2(\tau)\vf_2(z)}\,\sigma_2+
\frac{\nu^\pm_3}{c_3(\tau)\vf_3(z)}\,\sigma_3
  }
 \end{array}
\eq
 with the set of theta-constants
 \beq\label{e491}
 \begin{array}{c}
  \displaystyle{
c_k(\tau)=\Big(\frac{\theta_{5-k}(0)}{\theta_1'(0)}\Big)^2\,,\quad k=1,2,3\,.
  }
 \end{array}
\eq
 Each $K$-matrix $K^+(z)$ and $K^-(z)$ depends on four coupling constants $\nu_i^+$ and $\nu^-_i$, $i=0,1,2,3$ respectively.
Equivalently, expressions (\ref{e49}) are represented as follows:
 \beq\label{e50}
 \begin{array}{c}
  \displaystyle{
K^\pm(z)=\sigma_0\nu^\pm_0+\sigma_1\nu^\pm_1\vf_1(z+\om_1)
-\sigma_2\nu^\pm_2\vf_2(z+\om_2)-\sigma_3\nu^\pm_3\vf_3(z+\om_3)\,.
  }
 \end{array}
\eq
\noindent {\bf Proposition}
{\em
The gauge transformed $K$-matrices ${\ti K}(z)$ (\ref{e48}) defined through $K^\pm(z)$ (\ref{e49})
with the gauge transformation (\ref{e02}) take the following form:
 \beq\label{e52}
 \begin{array}{c}
  \displaystyle{
{\ti K}^+(z)=\mat{\nu^+_0}{0}{0}{\nu^+_0}+\sum\limits_{k=1}^3
\nu^+_k\mat{-\vf_k(2\bfq_1)}{\vf_k(2\bfq_1,z+\om_k)}{\vf_k(-2\bfq_1,z+\om_k)}{\vf_k(2\bfq_1)}\,,
  }
  \\ \ \\
  \displaystyle{
{\ti K}^-(z)=\mat{\nu^-_0}{0}{0}{\nu^-_0}+\sum\limits_{k=1}^3
\nu^-_k\mat{-\vf_k(2\bfq_n)}{\vf_k(2\bfq_n,z+\om_k)}{\vf_k(-2\bfq_n,z+\om_k)}{\vf_k(2\bfq_n)}\,,
  }
 \end{array}
\eq
 where the set of functions
 \beq\label{e521}
 \begin{array}{c}
  \displaystyle{
\vf_{k}(z,x+\om_k)=\frac{\theta_1'(0)\theta_{5-k}(z+x)}{\theta_{1}(z)\theta_{5-k}(x)}\,,\quad k=1,2,3\,,
  }
 \end{array}
\eq
 \beq\label{e522}
 \begin{array}{c}
  \displaystyle{
\vf_k(z,x+\om_k)\big|_{x=0}=\vf_k(z)\,.
  }
 \end{array}
\eq
 is used.
}

The proof is by direct computation
using identities with doubled modular parameter $2\tau$:
\beq\label{e51}
\begin{array}{c}
   \displaystyle{
\theta_2(x+y|2\tau)\theta_2(x-y|2\tau)=\frac{1}{2}\Big( \theta_3(x|\tau)\theta_3(y|\tau)-
\theta_4(x|\tau)\theta_4(y|\tau) \Big)\,,
 }
 \\ \ \\
    \displaystyle{
\theta_2(x+y|2\tau)\theta_3(x-y|2\tau)=\frac{1}{2}\Big( \theta_2(x|\tau)\theta_2(y|\tau)-
\theta_1(x|\tau)\theta_1(y|\tau) \Big)\,,
 }
  \\ \ \\
   \displaystyle{
\theta_3(x+y|2\tau)\theta_3(x-y|2\tau)=\frac{1}{2}\Big( \theta_3(x|\tau)\theta_3(y|\tau)+
\theta_4(x|\tau)\theta_4(y|\tau) \Big)\,.
 }
\end{array}
\eq
In particular, it provides $\det g(z,\bfq_a)=-\theta_1(z)/\theta_1(2\bfq_a)$.

Let us remark that similar calculation was performed for relation between ${\rm BC}_1$ Calogero-Inozemtsev model and the
non-relativistic Zhukovsky-Volterra gyrostat in \cite{LOZ05} as well as for relation between ${\rm BC}_1$ Ruijsenaars-van Diejen model and the
relativistic Zhukovsky-Volterra gyrostat in recent paper \cite{MoZ}.

 \section{Open elliptic Toda chain with boundaries}\label{sec6}
 By construction,
\beq\label{e60}
  \begin{array}{c}
  \displaystyle{
    \{ \tr{\mathcal T}^{\hbox{\tiny{openToda}}}(z), \tr{\mathcal T}^{\hbox{\tiny{openToda}}}(w) \}=
  \{ \tr{\mathcal T}^{\hbox{\tiny{openXYZ}}}(z), \tr{\mathcal T}^{\hbox{\tiny{openXYZ}}}(w) \}=0\,.
  }
 \end{array}
 \eq
 Let us compute the Hamiltonian for the open elliptic Toda chain
 generated by $\tr{\mathcal T}^{\hbox{\tiny{openToda}}}(z)$ (\ref{e47}).

 \paragraph{Transfer-matrix.} Consider the coefficient behind $1/z^{2n-2}$ in the expansion of
 $\tr{\mathcal T}^{\hbox{\tiny{openToda}}}(z)$ near $z=0$:
\beq\label{e61}
  \begin{array}{c}
  \displaystyle{
    H=\res\limits_{z=0}z^{2n-3}\tr{\mathcal T}^{\hbox{\tiny{openToda}}}(z)\,.
  }
 \end{array}
 \eq
Notice that matrices $L^{[a,b]}(z)$ (\ref{e04}) have only simple poles at $z=0$
due to
\beq\label{e611}
  \begin{array}{c}
  \displaystyle{
    \res\limits_{z=0}\phi(z,u)=1
  }
 \end{array}
 \eq
for the function $\phi(z,u)$ (\ref{e041}). The product $\tr{\mathcal T}^{\hbox{\tiny{openToda}}}(z)$ (\ref{e47})
contains $2n-2$ matrices $L^{[a,b]}(z)$, and the boundary $K$-matrices ${\ti K}^\pm(z)$ (\ref{e52}) are regular at
$z=0$.
Therefore,
\beq\label{e62}
  \begin{array}{c}
  \displaystyle{
    H=
\tr\Big({\ti K}^+(0)e^{P^1/2c}A^{[1,2]}\dots A^{[n-1,n]}{\ti K}^-(0)e^{P^n/2c}A^{[n,n-1]}\dots A^{[2,1]}\Big)
  }
 \end{array}
 \eq
with
\beq\label{e621}
  \begin{array}{c}
  \displaystyle{
A^{[k,m]}=\res\limits_{z=0}L^{[k,m]}(z)\,.
  }
 \end{array}
 \eq
The residues $A^{[k,m]}$ are easily calculated using the property (\ref{e611}):
\beq\label{e63}
  \begin{array}{c}
  \displaystyle{
A^{[k,m]}=\mats{b_1^{[k,m]}}{b_2^{[k,m]}}{b_1^{[k,m]}}{b_2^{[k,m]}}=
\left(\begin{array}{c}
1\\ 1
\end{array}\right)
\otimes (b_1^{[k,m]}\,, b_2^{[k,m]})\,,
  }
 \end{array}
 \eq
where $(b_1^{[k,m]}\,, b_2^{[k,m]})$ are 2-dimensional row-vectors.

 \paragraph{Expressions for ${\ti K}^\pm(0)$.} Taking into account the oddness (\ref{e4311}) of $\vf_k(z)$ and the definitions (\ref{e521})-(\ref{e522}), from (\ref{e52})
  we have
 \beq\label{e640}
 \begin{array}{c}
  \displaystyle{
{\ti K}^+(0)=\mat{\nu^+_0}{0}{0}{\nu^+_0}+\sum\limits_{k=1}^3
\nu^+_k\mat{-\vf_k(2\bfq_1)}{\vf_k(2\bfq_1)}{-\vf_k(2\bfq_1)}{\vf_k(2\bfq_1)}\,,
  }
  \\ \ \\
  \displaystyle{
{\ti K}^-(0)=\mat{\nu^-_0}{0}{0}{\nu^-_0}+\sum\limits_{k=1}^3
\nu^-_k\mat{-\vf_k(2\bfq_n)}{\vf_k(2\bfq_n)}{-\vf_k(2\bfq_n)}{\vf_k(2\bfq_n)}\,,
  }
 \end{array}
\eq
 Put it differently,
\beq\label{e64}
  \begin{array}{c}
  \displaystyle{
{\ti K}^+(0)=\mats{\nu^+_0}{0}{0}{\nu^+_0}-{\ti f}^+(\bfq_1)
\left(\begin{array}{c}
1\\ 1
\end{array}\right)
\otimes (1\,, -1)
  }
 \end{array}
 \eq
with
\beq\label{e641}
  \begin{array}{c}
  \displaystyle{
{\ti f}^+(\bfq_1)=\sum\limits_{k=1}^3\nu^+_k\vf_k(2\bfq_1)
  }
 \end{array}
 \eq
and similarly
%
\beq\label{e65}
  \begin{array}{c}
  \displaystyle{
{\ti K}^-(0)=\mats{\nu^-_0}{0}{0}{\nu^-_0}-{\ti f}^-(\bfq_n)
\left(\begin{array}{c}
1\\ 1
\end{array}\right)
\otimes (1\,, -1)\,,
  }
 \end{array}
 \eq
\beq\label{e651}
  \begin{array}{c}
  \displaystyle{
{\ti f}^-(\bfq_n)=\sum\limits_{k=1}^3\nu^-_k\vf_k(2\bfq_n)\,.
  }
 \end{array}
 \eq
One may use an alternative representation for the functions ${\ti f}^+(\bfq_1)$ and  ${\ti f}^-(\bfq_n)$:
\beq\label{e67}
  \begin{array}{c}
  \displaystyle{
\big({\ti f}^+(\bfq_1)\big)^2=\sum\limits_{k=0}^3({\bar\nu}_k^+)^2\wp(\bfq_1+\om_k)
-\sum\limits_{k=1}^3({\nu}_k^+)^2\wp(\om_k)\,,
  }
  \\ \ \\
    \displaystyle{
\big({\ti f}^-(\bfq_n)\big)^2=\sum\limits_{k=0}^3({\bar\nu}_k^-)^2\wp(\bfq_n+\om_k)
-\sum\limits_{k=1}^3({\nu}_k^-)^2\wp(\om_k)\,,
  }
 \end{array}
 \eq
where
  \beq\label{e68}
 \begin{array}{c}
  \left(\begin{array}{c}
 \bar\nu^\pm_0
 \\
 \bar\nu^\pm_1
 \\
 \bar\nu^\pm_2
 \\
 \bar\nu^\pm_3
 \end{array}\right)
 =
   \displaystyle{\frac12}
 \left(\begin{array}{ccc}
  1 & 1 & 1
 \\
   1 & -1 & -1
 \\
   -1 & 1 & -1
 \\
   -1 & -1 & 1
 \end{array}\right)
  \left(\begin{array}{c}
 \nu^\pm_1
 \\
 \nu^\pm_2
 \\
 \nu^\pm_3
 \end{array}\right)\,.
 \end{array}
 \eq
 This type identities can be found in \cite{MoZ}.

 \paragraph{The Hamiltonian.}
 Let us compute the expression for $H$ (\ref{e62}). Plugging (\ref{e63}) into (\ref{e62}) we get:
\beq\label{e652}
  \begin{array}{c}
  \displaystyle{
H=\prod\limits_{a=2}^{n-1}\Big(b_1^{[a-1,a]}+ b_2^{[a-1,a]}\Big)\Big(b_1^{[a+1,a]}+ b_2^{[a+1,a]}\Big)\times
  }
  \\ \ \\
  \displaystyle{
  \times \left[ \big(b_1^{[2,1]}\,, b_2^{[2,1]}\big){\ti K}^+(0)e^{P^1/2c}
\left(\begin{array}{c}
1\\ 1
\end{array}\right) \right]
 \times
 \left[ \big(b_1^{[n-1,n]}\,, b_2^{[n-1,n]}\big){\ti K}^-(0)e^{P^n/2c}
\left(\begin{array}{c}
1\\ 1
\end{array}\right) \right]\,.
  }
 \end{array}
 \eq
It follows from (\ref{e06}) that
\beq\label{e653}
  \begin{array}{c}
  \displaystyle{
b_1^{[a-1,a]}+ b_2^{[a-1,a]}=2\frac{\theta_1(\bfq_a-\bfq_{a-1})\theta_1(\bfq_a+\bfq_{a-1})}{\theta_1'(0)\theta_1(2\bfq_a)}
\sinh\big(\frac{\bfp_a}{2c}\big)\,,
  }
  \\ \ \\
  \displaystyle{
b_1^{[a+1,a]}+ b_2^{[a+1,a]}=2\frac{\theta_1(\bfq_a-\bfq_{a+1})\theta_1(\bfq_a+\bfq_{a+1})}{\theta_1'(0)\theta_1(2\bfq_a)}
\sinh\big(\frac{\bfp_a}{2c}\big)\,.
  }
 \end{array}
 \eq
Then for the upper line of (\ref{e652}) one obtains:
\beq\label{e654}
  \begin{array}{c}
   \displaystyle{
\prod\limits_{a=2}^{n-1}\Big(b_1^{[a-1,a]}+ b_2^{[a-1,a]}\Big)\Big(b_1^{[a+1,a]}+ b_2^{[a+1,a]}\Big)=
  }
  \\ \ \\
  \displaystyle{
=
\theta_1(\bfq_2-\bfq_{1})\theta_1(\bfq_2+\bfq_{1})\theta_1(\bfq_{n-1}-\bfq_{n})\theta_1(\bfq_{n-1}+\bfq_{n})\times
  }
  \\ \ \\
  \displaystyle{
\times(-1)^{n-3}\Big(\frac{2}{\theta_1'(0)}\Big)^{2n-4}\,
\left(\frac{\displaystyle{ \prod\limits_{a=3}^{n-1}\theta_1(\bfq_a-\bfq_{a-1})\theta_1(\bfq_a+\bfq_{a-1}) }}
{\displaystyle{ \prod\limits_{a=2}^{n-1} \theta_1(2\bfq_a)}}\right)^2\,\prod\limits_{a=2}^{n-1}\sinh^2\big(\frac{\bfp_a}{2c}\big)\,.
  }
 \end{array}
 \eq
Consider the lower line of (\ref{e652}). Using (\ref{e06}) and (\ref{e64}) we have
\beq\label{e655}
  \begin{array}{c}
  \displaystyle{
\big(b_1^{[2,1]}\,, b_2^{[2,1]}\big){\ti K}^+(0)e^{P^1/2c}
\left(\begin{array}{c}
1\\ 1
\end{array}\right) =
  }
  \\ \ \\
  \displaystyle{
=\frac{\theta_1(\bfq_1-\bfq_{2})\theta_1(\bfq_1+\bfq_{2})}{\theta_1'(0)\theta_1(2\bfq_1)}
\big(e^{\bfp_1/2c}\,, -e^{-\bfp_1/2c}\big){\ti K}^+(0)
\left(\begin{array}{c}
e^{\bfp_1/2c}\\ e^{-\bfp_1/2c}
\end{array}\right)=
  }
    \\ \ \\
  \displaystyle{
=\frac{\theta_1(\bfq_1-\bfq_{2})\theta_1(\bfq_1+\bfq_{2})}{\theta_1'(0)\theta_1(2\bfq_1)}
\Big(2\nu_0^+\sinh\big(\frac{\bfp_1}{c}\big)-4\sinh^2\big(\frac{\bfp_1}{2c}\big){\ti f}^+(\bfq_1)\Big)=
  }
      \\ \ \\
  \displaystyle{
=4\sinh\big(\frac{\bfp_1}{2c}\big)\frac{\theta_1(\bfq_1-\bfq_{2})\theta_1(\bfq_1+\bfq_{2})}{\theta_1'(0)\theta_1(2\bfq_1)}
\Big(\nu_0^+\cosh\big(\frac{\bfp_1}{2c}\big)-\sinh\big(\frac{\bfp_1}{2c}\big){\ti f}^+(\bfq_1)\Big)=
  }
        \\ \ \\
  \displaystyle{
=4\sinh^2\big(\frac{\bfp_1}{2c}\big)\frac{\theta_1(\bfq_1-\bfq_{2})\theta_1(\bfq_1+\bfq_{2})}{\theta_1'(0)\theta_1(2\bfq_1)}
\Big(\nu_0^+\coth\big(\frac{\bfp_1}{2c}\big)-{\ti f}^+(\bfq_1)\Big)\,.
  }
 \end{array}
 \eq
Similarly,
\beq\label{e656}
  \begin{array}{c}
  \displaystyle{
\big(b_1^{[n-1,n]}\,, b_2^{[n-1,n]}\big){\ti K}^-(0)e^{P^n/2c}
\left(\begin{array}{c}
1\\ 1
\end{array}\right) =
  }
      \\ \ \\
  \displaystyle{
=4\sinh\big(\frac{\bfp_n}{2c}\big)\frac{\theta_1(\bfq_n-\bfq_{n-1})\theta_1(\bfq_n+\bfq_{n-1})}{\theta_1'(0)\theta_1(2\bfq_n)}
\Big(\nu_0^-\cosh\big(\frac{\bfp_n}{2c}\big)-\sinh\big(\frac{\bfp_n}{2c}\big){\ti f}^-(\bfq_n)\Big)=
  }
        \\ \ \\
  \displaystyle{
=4\sinh^2\big(\frac{\bfp_n}{2c}\big)\frac{\theta_1(\bfq_n-\bfq_{n-1})\theta_1(\bfq_n+\bfq_{n-1})}{\theta_1'(0)\theta_1(2\bfq_n)}
\Big(\nu_0^-\coth\big(\frac{\bfp_n}{2c}\big)-{\ti f}^-(\bfq_n)\Big)\,.
  }
 \end{array}
 \eq
Plugging (\ref{e654}), (\ref{e655}) and (\ref{e656}) into (\ref{e652}) we come to the following expression for $H$:
\beq\label{e657}
  \begin{array}{c}
  \displaystyle{
H=
\frac{4^n}{\theta_1'(0)^{2n-2}}\,\frac{(-1)^{n-1}}{\theta_1(2\bfq_1)\theta_1(2\bfq_n)}\,
\left(\frac{\displaystyle{ \prod\limits_{a=2}^{n}\theta_1(\bfq_a-\bfq_{a-1})\theta_1(\bfq_a+\bfq_{a-1}) }}
{\displaystyle{ \prod\limits_{a=2}^{n-1} \theta_1(2\bfq_a)}}\right)^2\,\prod\limits_{a=1}^{n}\sinh^2\big(\frac{\bfp_a}{2c}\big)\times
  }
  \\ \ \\
  \displaystyle{
  \times
  \Big(\nu_0^+\coth\big(\frac{\bfp_1}{2c}\big)-{\ti f}^+(\bfq_1)\Big)
  \Big(\nu_0^-\coth\big(\frac{\bfp_n}{2c}\big)-{\ti f}^-(\bfq_n)\Big)\,.
  }
 \end{array}
 \eq
Next, let us once more use the identity (\ref{e101}):
\beq\label{e658}
  \begin{array}{c}
  \displaystyle{
\wp(\bfq_a-\bfq_{a-1})-\wp(\bfq_a+\bfq_{a-1})=
-\phi(\bfq_a-\bfq_{a-1},\bfq_a+\bfq_{a-1})\phi(\bfq_{a-1}-\bfq_{a},\bfq_a+\bfq_{a-1})=
}
\\ \ \\
  \displaystyle{
=\frac{\theta_1'(0)^2\theta_1(2\bfq_a)\theta_1(2\bfq_{a-1})}{\theta_1(\bfq_a-\bfq_{a-1})^2\theta_1(\bfq_a+\bfq_{a-1})^2}\,.
}
 \end{array}
 \eq
Then
\beq\label{e659}
  \begin{array}{c}
  \displaystyle{
\prod\limits_{a=2}^{n}\frac{1}{\wp(\bfq_a-\bfq_{a-1})-\wp(\bfq_a+\bfq_{a-1})}=
}
\\ \ \\
  \displaystyle{
=\frac{1}{\theta_1'(0)^{2n-2}}\,\frac{1}{\theta_1(2\bfq_1)\theta_1(2\bfq_n)}\,
\left(\frac{\displaystyle{ \prod\limits_{a=2}^{n}\theta_1(\bfq_a-\bfq_{a-1})\theta_1(\bfq_a+\bfq_{a-1}) }}
{\displaystyle{ \prod\limits_{a=2}^{n-1} \theta_1(2\bfq_a)}}\right)^2\,.
}
 \end{array}
 \eq
By comparing (\ref{e657}) and (\ref{e659}) one gets
\beq\label{e660}
  \begin{array}{c}
  \displaystyle{
H=(-1)^{n-1}4^n\,
\prod\limits_{a=2}^{n}\frac{1}{\wp(\bfq_a-\bfq_{a-1})-\wp(\bfq_a+\bfq_{a-1})}\,
\prod\limits_{a=1}^{n}\sinh^2\big(\frac{\bfp_a}{2c}\big)\times
  }
  \\ \ \\
  \displaystyle{
  \times
  \Big(\nu_0^+\coth\big(\frac{\bfp_1}{2c}\big)-{\ti f}^+(\bfq_1)\Big)
  \Big(\nu_0^-\coth\big(\frac{\bfp_n}{2c}\big)-{\ti f}^-(\bfq_n)\Big)\,.
  }
 \end{array}
 \eq
 Finally, we come to the following Hamiltonian for the open elliptic Toda chain with boundaries:
  \beq\label{e70}
  \begin{array}{c}
  \displaystyle{
H^{\hbox{\tiny{openToda}}}=-\log(H)+\log((-1)^{n-1}{4^n})=
}
 \\ \ \\
\displaystyle{
  =\sum\limits_{a=1}^{n}\log\frac{1}{\sinh^2(\bfp_a/2c)}
+
\sum\limits_{a=2}^n\log\Big(\wp(\bfq_{a-1}-\bfq_a)-\wp(\bfq_{a-1}+\bfq_a)\Big)
-
}
 \\ \ \\
\displaystyle{
-\log\Big(\nu_0^+\coth\big(\frac{\bfp_1}{2c}\big)-{\ti f}^+(\bfq_1)\Big)
-\log\Big(\nu_0^-\coth\big(\frac{\bfp_n}{2c}\big)-{\ti f}^-(\bfq_n)\Big)\,.
}
 \end{array}
 \eq
To summarize, we proved the following statement.

\noindent{\bf Theorem}
{\em
The Hamiltonian for the open elliptic Toda chain with boundaries evaluated from the classical
monodromy matrix ${\mathcal T}^{\hbox{\tiny{openToda}}}(z)$ (\ref{e47}) has the form (\ref{e70}),
where the functions ${\ti f}^+(\bfq_1)$, ${\ti f}^-(\bfq_n)$ are defined in (\ref{e641}), (\ref{e651}) or (\ref{e67}).
By construction, this model is gauge equivalent to the
open XYZ chain with boundaries given by $K$-matrices (\ref{e49}), and with
the Casimir functions fixed as in (\ref{e22}) at each site.
}

Compared to the original Hamiltonian (\ref{e01}) for the closed elliptic Toda chain, the expression (\ref{e70})
does not contain interaction of $n$-th particle with the first one. But it contains boundary terms
for these two particles on boundaries.

Consider two particular cases.

\noindent{\bf Example 1: pure open chain.} Let $\nu^\pm_k=0$, $k=1,2,3$ and $\nu^\pm_0=1$. From viewpoint
of the open XYZ chain this case
corresponds to trivial $K$-matrices $K^\pm(z)=\sigma_0$ (\ref{e49}), and the gauge transformed
$K$-matrices (\ref{e52}) are trivial as well: ${\ti K}^\pm(z)=\sigma_0$. Then
  \beq\label{e71}
  \begin{array}{c}
  \displaystyle{
H^{\hbox{\tiny{pure\ open}}}=
\log\frac{1}{\sinh(\bfp_1/2c)\cosh(\bfp_1/2c)}
+\log\frac{1}{\sinh(\bfp_n/2c)\cosh(\bfp_n/2c)}+
}
 \\ \ \\
\displaystyle{
  +\sum\limits_{a=2}^{n-1}\log\frac{1}{\sinh^2(\bfp_a/2c)}
+
\sum\limits_{a=2}^n\log\Big(\wp(\bfq_{a-1}-\bfq_a)-\wp(\bfq_{a-1}+\bfq_a)\Big)\,.
 }
 \end{array}
 \eq
This Hamiltonian differs from (\ref{e01}) by factor $-1/2$, by the
absence of interaction between the first and the last particles, and by
type of dependence on momenta for these two particles on the boundaries.

\noindent{\bf Example 2: only boundary terms.} Now let it be
vice versa: $\nu^\pm_0=0$. Using (\ref{e67}) we obtain
  \beq\label{e72}
  \begin{array}{c}
  \displaystyle{
H^{\hbox{\tiny{boundary}}}=\sum\limits_{a=1}^{n}\log\frac{1}{\sinh^2(\bfp_a/2c)}
+
\sum\limits_{a=2}^n\log\Big(\wp(\bfq_{a-1}-\bfq_a)-\wp(\bfq_{a-1}+\bfq_a)\Big)
-
}
 \end{array}
 \eq
$$
\displaystyle{
-\frac12\log\Big( \sum\limits_{k=0}^3({\bar\nu}_k^+)^2\wp(\bfq_1+\om_k)
-\sum\limits_{k=1}^3({\nu}_k^+)^2\wp(\om_k) \Big)
-\frac12\log\Big( \sum\limits_{k=0}^3({\bar\nu}_k^-)^2\wp(\bfq_n+\om_k)
-\sum\limits_{k=1}^3({\nu}_k^-)^2\wp(\om_k) \Big)\,.
}
$$
In this case the kinetic part (dependence on momenta) is the same as in (\ref{e01}) (up to factor $-1/2$) but there are
additional external fields for the first and the last particles on the boundaries.

In conclusion, we notice again that the Ruijsenaars-Toda chain \cite{Adler} generalizes
the elliptic Toda chain (\ref{e01}), and it is gauge equivalent
to a generic XYZ chain (without restrictions (\ref{e22})). Similarly to the results of the present paper
it is then possible to derive the open Ruijsenaars-Toda chain with boundaries. The boundary
$K$-matrices can be also chosen to be dynamical, see the Zhukovsky-Volterra gyrostat \cite{LOZ05} and
the 4-constant ${\rm BC}_1$ Ruijsenaars-van Diejen model \cite{MoZ}. These generalizations will be studied
elsewhere.








\paragraph{Acknowledgments.} We are grateful to D. Murinov and A. Zabrodin for discussions.

This work was supported by the Russian Science Foundation under grant no. 25-11-00081, \\ 
https://rscf.ru/project/25-11-00081/ .


\begin{footnotesize}

\end{footnotesize}


\begin{thebibliography}{99}
\addcontentsline{toc}{section}{References}



\bibitem{AdlerShabat}
 V.E. Adler, A.B. Shabat,
  {\it On a class of Toda chains},
  Theoret. and Math. Phys. 111:3 (1997),
647--657.

\bibitem{Adler} V.E. Adler, Y.B. Suris,
{\it Q$_4$: integrable master equation related to an elliptic curve},
International Mathematics Research Notices, vol. 2004:47 (2004) 2523--2553.

\bibitem{AtZ} K. Atalikov, A. Zotov, {\it Field theory generalizations of two-body Calogero-Moser models in the form of Landau-Lifshitz equations}, 	J. Geom. Phys., 164 (2021) 104161; 	arXiv:2010.14297 [hep-th].

\bibitem{Baxter2} R.J. Baxter,
 {\it Eight-vertex model in lattice statistics and
one-dimensional anisotropic Heisenberg chain. II. Equivalence to a
generalized ice-type lattice model},
Ann. Phys. 76 (1973) 25--47.

\bibitem{FT} L.D. Faddeev, L.A. Takhtajan,
{\it Hamiltonian methods in the theory of solitons},
Berlin; New York:
Springer-Verlag (1987).

\bibitem{Has}
K. Hasegawa, {\it Ruijsenaars' Commuting Difference Operators as
Commuting Transfer Matrices}, Commun. Math. Phys. 187 (1997)
289--325,    arXiv:q-alg/9512029.

\bibitem{IK} T. Inami, H. Konno,
{\it Integrable XYZ spin chain with boundaries},
J. Phys. A Math. Gen. 27 (1994) L913--L918.

Y. Komori, K. Hikami,
 {\it Elliptic K-matrix associated with Belavin's symmetric R-matrix},
 Nuclear Physics B 494:3 (1997) 687-701.

\bibitem{KrToda} I. Krichever,
{\it Elliptic analog of the Toda lattice},
International Mathematics Research Notices, vol. 2000, no. 8 (2000) 383--412;
	arXiv:hep-th/9909224.

\bibitem{LOZ} A. Levin, M. Olshanetsky, A. Zotov,
 {\it Hitchin Systems -- Symplectic Hecke Correspondence and Two-dimensional Version},
Commun. Math. Phys.  236 (2003) 93--133;     arXiv:nlin/0110045.



\bibitem{LOZ05}  A.M. Levin, M.A. Olshanetsky, A.V. Zotov,
 {\it Painlev\'e VI, rigid tops and reflection equation},
Commun. Math. Phys., 268:1 (2006) 67--103;
 	arXiv:math/0508058
  [math.QA].

\bibitem{MoZ} A.M. Mostovskii, A.V. Zotov,
 {\it Classical elliptic ${\rm BC_1}$
 Ruijsenaars–van Diejen model: relation to Zhukovsky–Volterra gyrostat and 1-site classical XYZ
 model with boundaries},
 Theoret. and Math. Phys., 226:2 (2026) 189--216;
 	arXiv:2601.06826 [math-ph].

\bibitem{MuZ1} D. Murinov, A. Zotov,
{\it Classical r-matrix structure for elliptic Ruijsenaars chain and 1+1
 field analogue of Ruijsenaars–Schneider model},
  J. Phys. A: Math. Theor., 58:50 (2025) 505205; 	arXiv:2508.12656 [math-ph].


\bibitem{MuZ2} D. Murinov, A. Zotov,
{\it Elliptic Ruijsenaars-Toda and elliptic Toda chains: classical r-matrix structure and relation to XYZ chain},
Journal of Geometry and Physics, 226 (2026), 105848;
arXiv:2602.08143 [nlin.SI].

\bibitem{Skl2}
E.K. Sklyanin,
{\it Some algebraic structures connected with the Yang-Baxter equation},
Funct. Anal. Appl. 16 (1982) 263--270.

E.K. Sklyanin,
{\it Some algebraic structures connected with the Yang-Baxter equation. Representations of quantum algebras},
Funct. Anal. Appl. 17 (1983) 273--284.

\bibitem{Skl4}
E.K. Sklyanin,
 {\it On the Poisson structure of the periodic classical XYZ-chain},
 Questions of quantum field theory and statistical physics. Part 6,
 Zap. Nauchn. Sem. LOMI, 150, (1986) 154--180;
{\it Poisson structure of a periodic classical XYZ-chain}, J. Sov. Math. 46 (1989) 1664.

\bibitem{Skl-refl} E.K. Sklyanin,
 {\it Boundary conditions for integrable equations},
 Funct. Anal. Appl., 21:2 (1987) 164--166.

E. Sklyanin,
{\it Boundary conditions for integrable quantum systems},
J. Phys. A: Math. Gen. 21 (1988) 2375--2389.

\bibitem{Suris} Yuri B. Suris,
 {\it Discrete time Toda systems},
 J. Phys. A: Math. Theor. 51 (2018) 333001.

\bibitem{VZ} M. Vasilyev, A. Zotov,
 {\em On factorized Lax pairs for classical many-body integrable systems},
Reviews in Mathematical Physics, 31:6 (2019) 1930002; 	arXiv:1804.02777 [math-ph].

\bibitem{Yamilov} R.I. Yamilov, {\it Generalizations of the Toda chain, and conservation laws}, Preprint Inst. of
Math., Ufa (1989) (in Russian). English version: {\it Classification of Toda–type scalar lattices}
in  {\it Nonlinear evolution equations and dynamical systems}, NEEDS’92, Eds. V. Makhankov,
I. Puzynin, O. Pashaev (1993). Singapore: World Scientific, 423--431.

\bibitem{ZZ} A. Zabrodin, A. Zotov,
{\it Field analogue of the Ruijsenaars-Schneider model},
JHEP 07 (2022) 023;	arXiv: 2107.01697 [math-ph].


\end{thebibliography}
\end{document}